# Axial and polar magnetism in hexagonal YMnO$_3$

S. W. Lovesey[1, 2, 3]

[1] ISIS Facility, STFC, Didcot, Oxfordshire OX11 0QX, UK

[2] Diamond Light Source, Harwell Science and Innovation Campus, Didcot, Oxfordshire OX11 0DE, UK

[3] Department of Physics, Oxford University, Oxford OX1 3PU, UK

**Abstract**. Newly published diffraction data on hexagonal YMnO$_3$ at a temperature of 10 K are shown to be consistent with a trusted expression of the magnetic symmetry, although the data alone are not definitive [M. Ramakrishnan *et al.*, Phys. Rev. Research **5**, 013203 (2023)]. Howard *et al.* conclude from an exhaustive review of experimental data that the symmetry of the antiferromagnetic motif of Mn ions is most likely P6$_3$'cm' [C. J. Howard *et al.*, Acta Crystallogr. B **69**, 534 (2013)]. The data reported by Ramakrishnan *et al.* does not eliminate symmetry P6$_3$' from our calculated diffraction patterns, because the studied reflection vector is parallel to the common chiral vector associated with each triangle of Mn axial dipole moments. Proposed diffraction patterns will give decisive statements about the magnetic symmetry in future investigations using resonant x-ray and magnetic neutron diffraction. To this end, both axial and polar magnetism in the multiferroic are essential in the analysis of diffraction patterns. The account by Ramakrishnan *et al.* of polar magnetism in resonant x-ray diffraction uses a magnetic symmetry of hexagonal YMnO$_3$ not yet justified, e.g., ferromagnetism and a linear magnetoelectric effect allowed by their chosen symmetry have not been observed. We study polar magnetism in symmetry P6$_3$'cm' using Dirac multipoles, including Mn anapoles. They also feature in amplitudes for magnetic neutron diffraction together with Dirac quadrupoles, previously shown to account for diffraction by pseudo-gap phases of cuprate superconductors.

## I. INTRODUCTION

Electronic and magnetic properties of hexagonal YMnO$_3$ are topics of many experimental and theoretical investigations. One reason being that it belongs to the rare class of multiferroic materials which exhibit both ferroelectricity and (noncollinear) magnetic order. The manganese trioxide is paraelectric at elevated temperatures and undergoes a single structural transition at ≈ 1250 K from a (centrosymmetric) to a (non-centrosymmetric) ferrielectric phase that is retained through room temperature [1]. The hexagonal structure hosts five- and sevenfold coordination polyhedra about Mn (Mn$^{3+}$) and Y (Y$^{3+}$) ions, respectively. Ferroelectricity might arise from buckling of the MnO$_5$ bipyramids [2]. A magnetic transition at T$_N$ ≈ 70 K heralds a triangular antiferromagnetic arrangement of Mn ions with a propagation vector **k** = 0 [3, 4, 5]. In addition to aforementioned properties, the manganite displays a giant magneto-electric coupling [6, 7].

By and large, structural and magnetic properties of hex-YMnO$_3$ have been inferred from Bragg diffraction experiments, although Fiebig *et al.* [3] demonstrate the value of second harmonic generation in determining magnetic symmetry. In **k** = 0 structures, magnetic reflections arise below $T_N$ at the same positions as the reflections of structural origin, e.g., the magnetic scattering of neutrons in general coincides with nuclear reflections. There are only a few pure magnetic reflections where structural reflections are not allowed due the space group extinctions. The refinement of the crystal and magnetic structures using overlapping nuclear and magnetic intensities at the same reciprocal lattice positions has the problem of separating magnetic and nuclear intensities and therefore leads to correlations between the magnetic and structural parameters. Likewise with x-ray diffraction, with overlapping Thomson and magnetic intensities. Not withstanding the mentioned limitations, there is compelling evidence that symmetry P6$_3$'cm' correctly expresses the ordered magnetic structure of hex-YMnO$_3$ [5]. (Howard *et al.* critically review the existence of an intermediate phase in the transition from the paraelectric structure to the ferroelectric structure [2, 5].) Another contender for the ordered magnetic structure of hex-YMnO$_3$ with a reduced symmetry, P6$_3$', is also surveyed [4, 5].

Bragg diffraction patterns for P6$_3$'cm' have not been published, to the best of our knowledge. Analytic calculations reported in this communication are informed by symmetry, and serve both neutron and x-ray diffraction. Enhancement of Bragg spot intensities obtained in resonant x-ray diffraction facilitates investigations of nominally weak Bragg spots not of structural origin, i.e., basis-forbidden reflections. We present diffraction patterns that include basis-forbidden reflections for charge-like, polar and Dirac (polar and magnetic) atomic entities. Many resonant x-ray diffraction studies using manganese K and L edges have been published [8-13]. Manganese L edges are used in recent experiments on hex-YMnO$_3$ [14]. Resonant enhancement brought about a relatively strong Bragg spot for a reflection condition forbidden in the parent structure P6$_3$cm (No. 185). Estimates of charge-like, polar and Dirac Mn multipoles, derived from *ab initio* simulations (FDMNES code) of an electronic structure using magnetic space group P6$_3$c'm' (No. 185.201), are also included [14, 15]. Howard *et al.* include symmetry P6$_3$'cm' used here, and P6$_3$c'm' in their group-subgroup relationships for the crystal and magnetic structures (Fig. 3), together with summaries of magnetic properties (Table 1) [5]. Notably, weak ferromagnetism and a magnetoelectric effect are allowed in P6$_3$c'm', and both these properties are forbidden by symmetry in P6$_3$'cm'.

Atomic Mn multipoles used in our diffraction amplitudes possess discrete symmetries of space and time. Polar varieties are permitted because Mn ions in hex-YMnO$_3$ occupy sites that do not possess a centre of spatial inversion. Polar multipoles (parity-odd and nonmagnetic) are allowed in diffraction amplitudes for polar crystal structures, and they diffract x-rays. Likewise, magnetic multipoles that are either parity-even (conventional axial magnetism) or parity-odd. Axial Mn dipoles in hex-YMnO$_3$ are depicted in Fig. 1 [16, 17]. The polar magnetic family, Dirac multipoles, include a magnetic monopole formed with (**S · R**), where **S** and **R** are electronic spin and position operators, respectively [18, 19, 20]. A generic Mn multipole $\langle O^K_Q \rangle$ has integer rank K, and angular brackets $\langle ... \rangle$ denote an expectation value. The (2K + 1) projections Q are in the interval $-K \leq Q \leq K$. Conditions on K and Q are imposed by both site and crystal symmetries, and the interaction of the radiation with electrons. Restrictions on K are imposed by the triangle rule in the case of resonant x-ray diffraction, e.g., K = 1, 2, 3 for enhancement by an electric dipole (E1) - electric quadrupole (E2) absorption event.

In the present case, site symmetry alone imposes the constraint that Mn axial multipoles are purely real for all K and Q. From this it follows that axial dipoles depicted in Fig. 1 are constrained to the (ac) plane. Likewise, manganese Dirac multipoles are purely real (imaginary) for even K (odd) for all Q, e.g., a magnetic monopole (K = Q = 0) is permitted. Projections Q are constrained by the crystal structure, however, when the reflection vector and the axis of rotation symmetry parallel to the c axis coincide. In this case, Q = 3$n$ where $n$ is an integer. Notably, multipoles with Q = ±3 are octupoles (K = 3) and those of higher order. A signature of symmetry P6$_3$'cm' is a condition on a Miller index, Q and the time signature of $\langle O^K_Q \rangle$. By way of an example of its consequences, axial magnetic dipoles are forbidden in bulk magnetism (hex-YMnO$_3$ is an antiferromagnet), while magnetic octupoles are permitted. An anapole depicted in Fig. 2 [18, 21], perhaps the second-best known Dirac multipole beyond the magnetic monopole, also contributes to diffraction patterns. All mentioned features of hex-YMnO$_3$ flow from an electronic structure factor Eq. (A1) for Mn ions in a unit cell with symmetry P6$_3$'cm'. With it, diffraction patterns of x-rays and neutrons can be calculated [20, 22, 23, 24]. Resonant x-ray diffraction and magnetic neutron diffraction are treated in the main text and Appendix C, respectively.

## II. MAGNETIC STRUCTURE

Vectors describing the hex-YMnO$_3$ unit cell in Fig. 1 are **a** = ($a$, 0, 0), **b** = (1/2) (−$a$, $a\sqrt{3}$, 0) and **c** = (0, 0, $c$) in an orthonormal coordinate system. Cell lengths $a$ ≈ 6.120 Å and $c$ ≈ 11.408 Å [6]. Local axes for Mn ions labelled (ξ, η, ζ) match orthogonal vectors **a**, **b**\* ∝ **a** + 2**b**, and **c**.

Manganese ions occupy sites 6c (≈ 0.342, 0, 0) in space group No. 185. 200 (P6$_3$'cm', BNS [16]) that possess anti-mirror symmetry m' along the tertiary symmetry direction η = [1, 2, 0] [5]. A Mn multipole $\langle O^K_Q \rangle$ obeys (σ$_\pi$ σ$_\theta$ 2$_\eta$) $\langle O^K_Q \rangle$ = [σ$_\pi$ σ$_\theta$ (−1)$^{K+Q}$] $\langle O^K_{-Q} \rangle$ = $\langle O^K_Q \rangle$, where σ$_\pi$ and σ$_\theta$ are signatures for parity and time, respectively [18, 20]. In a standard setting (−1)$^Q$ $\langle O^K_{-Q} \rangle$ = $\langle O^K_Q \rangle$*, where * denotes complex conjugation. Whence, diagonal (Q = 0) multipoles are purely real. The nominal Mn electronic configuration is 3d$^4$ (Mn$^{3+}$) with a high-spin configuration S = L = 2 and total angular momentum J = 0.

Magnetic symmetry P6$_3$'cm' belongs to the magnetic crystal class 6'mm' that is polar and compatible with the piezomagnetic effect. Ferromagnetism is not allowed. The Landau free-energy includes E, EHH and HEE, where E and H are electric and magnetic fields, respectively. The magnetic crystal class 6' possesses identical properties, and it is correct for symmetry P6$_3$' in Section IV.

X-ray and neutron diffraction amplitudes can be derived from an electronic structure factor,

$$\Psi^K_Q = [\exp(i\boldsymbol{\kappa} \cdot \mathbf{d}) \langle O^K_Q \rangle_\mathbf{d}], \qquad (1)$$

where the implied sum runs over the 6 Mn ions at sites **d** in a unit cell. The reflection vector $\boldsymbol{\kappa}$ = ($h$, $k$, $l$) with integer Miller indices. Manganese sites in a cell of hex-YMnO$_3$ are related by threefold rotations about the c axis, and anti-sixfold rotations about the c axis with translations **c**/2. Since rotations about the c axis do not change Q, the electronic structure factor is proportional to $\langle O^K_Q \rangle$ at the cell origin. A complete version of $\Psi^K_Q$ for sites 6c in P6$_3$'cm' is given in Eq. (A1).

Evaluated for a reflection vector $\boldsymbol{\kappa}$ = (0, 0, $l$) it reduces to,

$$\Psi^K_Q(6c) = \langle O^K_Q \rangle \, [1 + \sigma_\theta \, (-1)^{l+Q}] \, [1 + 2\cos(2\pi Q/3)], \qquad (2)$$

which is appropriate for experiments reported by Ramakrishnan *et al.* [14]. Later, we consider $\boldsymbol{\kappa}$ = ($h$, 0, $l$). The second bracket in Eq. (2) is different from zero for Q = 0, ±3, etc. Conditions imposed on Q are the mentioned consequence of the alignment of $\boldsymbol{\kappa}$ and a triad axis of rotation symmetry in the crystal. Evidently, crystal and magnetic symmetries are interrelated in a non-zero value of the first bracket. Nuclear (K = 0) and Thomson multipoles are time-even and $\sigma_\theta$ = +1. Axial multipoles ($\sigma_\pi$ = +1) in resonant x-ray diffraction have an even rank, but this restriction on K does not apply to Dirac multipoles ($\sigma_\pi$ = −1).

### III. RESONANT X-RAY DIFFRACTION

An atomic resonance in the x-ray absorption spectrum is often a sharp feature [13, 25]. In which case, it is meaningful to assign an amplitude to the resonant contribution equal to its energy-integrated intensity. The four amplitudes are labelled by polarization states depicted in Fig. 3, and they can be developed in electronic multipoles introduced in Section I [19, 20, 26]. Analytic expressions for axial and Dirac multipoles for an informative atomic model are listed by Lovesey and Scagnoli [18]. In our notation, ($\pi'\sigma$) denotes a rotated amplitude, and |($\pi'\sigma$)|$^2$ the intensity of the Bragg spot enhanced by the atomic resonance [18, 20]. Universal expressions for diffraction amplitudes employed here are functions of the rotation of the illuminated crystal about the reflection vector by an angle $\psi$ [26].

Parity-even (E1-E1, E2-E2) Mn multipoles denoted $\langle T^K_Q \rangle$ satisfy $(-1)^Q \langle T^K_{-Q} \rangle = \langle T^K_Q \rangle$, i.e., multipoles are purely real ($\sigma_\theta (-1)^K$ = +1, $\sigma_\pi$ = +1: E1-E1, K = 0-2; E2-E2, K = 0-4). Dirac multipoles $\langle G^K_Q \rangle$ possess discrete symmetries $\sigma_\pi \sigma_\theta$ = +1. Whence, E1-E2 multipoles are purely real (imaginary) for even K (odd K). Specifically, a magnetic charge (monopole $\langle G^0_0 \rangle$) that contributes in an E1-M1 event (K = 0-2) is allowed by site symmetry.

Returning to Eq. (2), we consider basis forbidden reflections with odd $l$. Conditions for non-zero $\Psi^K_Q$ are: E1-E1 & E2-E2; odd K + Q; E1-M1 & E1-E2; even K and Q = 0 for Dirac multipoles. Diagonal (Q = 0) contributions to (0, 0, $l$) x-ray diffraction amplitudes do not depend on the azimuthal angle.

Unrotated E1-E1 amplitudes are zero for $\kappa = (0, 0, l)$ with odd $l$. Whereas, $(\pi'\sigma)_{11}$ is proportional to the axial dipole moment along the crystal c axis, $\langle T^1_0 \rangle = \langle T^1_\zeta \rangle$. Specifically, $(\pi'\sigma)_{11}$ is purely imaginary with [26],

$$(\pi'\sigma)_{11} = (i3\sqrt{2}) \sin(\theta) \langle T^1_\zeta \rangle, \tag{3}$$

where $\theta$ is the Bragg angle depicted in Fig 3. An E2-E2 amplitude contains a hexadecapole; see below. Equal intensities are observed in Bragg diffraction by hex-YMnO$_3$ at Mn L$_2$ and L$_3$ edges [14]. This result implies a null orbital contribution from 3d (4p) electrons sampled in an E1-E1 (E2-E2) absorption event. The reasoning is as follows. Reduced matrix elements (RMEs) for both parity-even events are (A + B) and (2A − B) for L$_2$ and L$_3$ edges, respectively (e.g., [27] and Eq. (73) in Ref. [20]). Here, A is proportional to the orbital angular momentum in the 3d or 4p states. On the other hand, B includes expectation values of spin and hybrid spin-orbital operators. RMEs for K edges are independent of spin variables (B ≡ 0) [28]. More information on sum rules is given in Appendix B.

Unrotated Dirac E1-E2 amplitudes are zero while,

$$(\pi'\sigma)_{12} = (3/2\sqrt{5}) [1 - 3 \cos(2\theta)] \langle G^2_0 \rangle. \tag{4}$$

Note the 90° phase shift between E1-E1 and E1-E2 diffraction amplitudes. Intensities are in quadrature and interference between E1-E1 and E1-E2 intensities does not exist.

An anapole, illustrated in Fig. 2, $\langle G^1_\eta \rangle = -\sqrt{2} \langle G^1_{+1} \rangle''$ contributes to all $(h, 0, l)$ diffractions amplitudes, and we provide $(\sigma'\sigma)_{12}$ by way of an example. To this end, $r = a/c$, and

$$\sin(\beta) = rl\sqrt{3}/Z, \cos(\beta) = -2h/Z, \text{ with } Z = \sqrt{[4h^2 + (rl\sqrt{3})^2]}. \tag{5}$$

From the electronic structure factor Eq. (A1) evaluated for $(h, 0, l)$ and odd $l$,

$$(\sigma'\sigma)_{12} = i (2/5)\sqrt{2} \sin(2\pi hx) \cos(\theta) \cos(\psi) [3 \langle G^1_{+1} \rangle'' + \sqrt{5} \cos(2\beta) \langle G^2_{+1} \rangle'$$

$$+ 5 \{2 + 18 \cos^2(\beta) \cos^2(\psi) - 13 \cos^2(\beta) - 3 \cos^2(\psi)\} \langle G^3_{+1} \rangle'']. \tag{6}$$

The amplitude $(\pi'\sigma)_{12}$ in the rotated channel of polarization has contributions proportional to $[\sin(2\beta) \langle G^2_{+1} \rangle']$ and $[\sin(2\beta) \langle G^3_{+1} \rangle'']$ in addition. Reflections of the type $h = 3n$ are weak for hex-YMnO$_3$ since the general coordinate x is close to 1/3. When the azimuthal angle $\psi = 0$ the crystal axis **b** and $\eta$ coincide. The diffraction condition $(h, 0, l)$ with odd $l$ is not satisfied at Mn L edges (L$_2$ ≈ 0.649 keV and L$_3$ ≈ 0.638 keV), and it is at the K edge (≈ 6.537 keV) for a range of $h$ and $l$.

There is no diffraction by polar E1-E2 multipoles $\langle U^K_Q \rangle$ for odd $l$. Likewise for the rotated channel of polarization and (0, 0, $l$) with even $l$. Diagonal polar multipoles with odd K diffract in the unrotated channels, however. The dipole $\langle U^1_0 \rangle$ is the average displacement of the Mn ion along the crystal c axis. And for a reflection vector (0, 0, $l$) with even $l$,

$$(\sigma'\sigma)_{12} \propto i \sin(\theta) \, [\sqrt{3} \, \langle U^1_0 \rangle - \sqrt{2} \, \langle U^3_0 \rangle], \tag{6}$$

$$(\pi'\pi)_{12} \propto i \, [\sqrt{3} \sin(3\theta) \langle U^1_0 \rangle + \sqrt{2} \sin(\theta) \{1 + \cos^2(\theta)\} \langle U^3_0 \rangle].$$

Amplitudes do not depend on the azimuthal angle, as expected.

Returning to $\kappa$ = (0, 0, $l$) with odd $l$, the E2-E2 amplitude $(\pi'\sigma)_{22}$ possesses a signature of the triad axis of rotation symmetry along the c axis, in that $(\pi'\sigma)_{22}$ is three-fold periodic in the azimuthal angle $\psi$. An E2 absorption event at L edges uses electronic states 2p → 4p, and a hexadecapole $\langle T^4_{+3} \rangle'$ is not permitted. Specifically,

$$(\pi'\sigma)_{22} = (3/\sqrt{10}) \, [-i \, \{\sin(3\theta) \langle T^1_\zeta \rangle + \sin(\theta) (2 - 3\cos^2(\theta)) \langle T^3_0 \rangle\}$$

$$+ \sqrt{5} \cos^3(\theta) \cos(3\psi) \langle T^4_{+3} \rangle']. \tag{7}$$

The azimuthal angle scan starts with the crystal axis **a** normal to the plane of scattering. Notably, magnetic and charge-like contributions to $(\pi'\sigma)_{22}$ differ by a 90° phase and intensities are in quadrature.

### IV. REDUCED SYMMETRY

Howard *et al.* reject symmetry P6$_3$' for for hex-YMnO$_3$ for several reasons, and to further refine the debate about its relevance we survey diffraction properties of P6$_3$' (No. 173.131 BNS) [4, 5]. The c-glide plane perpendicular to [1, 0, 0] (and [0, 1, 0] and [−1, −1, 0]) and mirror plane perpendicular to [1, −1, 0] (and [1, 2, 0] and [−2, −1, 0]) are lost in P6$_3$', and the chiral vector parallel to the c axis associated with each triangle of Mn dipoles is all that is retained of P6$_3$'cm'. The reduced symmetry leaves Mn dipole moments at arbitrary angles to the crystallographic axes, because Mn ions in sites 6c are not constrained by any symmetry.

Scattering amplitudes for the reflection (0, 0, $l$) with odd $l$ are identical in P6$_3$' and P6$_3$'cm', and Eqs. (3) and (4) are valid with reduced symmetry. Not so for ($h$, 0, $l$) and odd $l$. We demonstrate distinguishing features of reduced symmetry by revisiting E1-E2 diffraction and the unrotated amplitude Eq. (6). At the level of anapoles and quadrupoles,

$$(\sigma'\sigma)_{12} = \text{Eq. (6)} + i \, (2/5)\sqrt{2} \sin(2\pi hx) \cos(\theta) \sin(\psi) \sin(\beta)$$

$$\times [3 \langle G^1_{+1} \rangle' + \sqrt{5} \langle G^2_{+1} \rangle'' + ... \,], \tag{8}$$

with $\langle G^1_\xi \rangle = -\sqrt{2} \langle G^1_{+1} \rangle'$, and β is defined in Eq. (5). Dirac multipoles forbidden in the higher symmetry P6$_3$'cm' have a different dependence on the azimuthal angle and β.

## V. DISCUSSION AND CONCLUSIONS

In summary, we have studied a space group (P6$_3$'cm') that is a well-founded expression of the magnetic structure of hex-YMnO$_3$ [5]. New diffraction data provides a motivation to add to the extensive literature on the multiferroic material [14]. Curiously, the diffraction pattern of P6$_3$'cm' has not been published. Here, patterns for magnetic neutron and resonant x-ray diffraction are derived from the electronic structure factor Eq. (A1) for Mn ions in sites 6c, which hosts axial and polar (Dirac) magnetism. We exploit elementary magnetic crystallography, whereby symmetries of occupied sites and the magnetic unit cell are paramount. Encapsulating atomic properties of a material in electronic multipoles with discrete symmetries, in both space and time, enables symmetry informed calculations of diffraction patterns that can be confronted with observations.

The chiral vector in P6$_3$'cm' parallel to the c axis associated with each triangle of Mn axial dipoles (Fig. 1) dictates the diffraction pattern for reflection vectors (0, 0, *l*). Specifically, it fixes angular anisotropy labelled by projections Q of the atomic multipoles $\langle T^K_Q \rangle$ with integer rank K and − K ≤ Q ≤ K. The result Q = 3*n* is self-evident, and scattering amplitudes for x-ray diffraction enhanced by an electric dipole - electric dipole (E1-E1, K = 0, 1, 2) absorption event use Q = 0. In consequence, Bragg spot intensities do not change with rotation of the crystal about the reflection vector (an azimuthal angle scan). We find no intensity in unrotated channels of polarization (Fig. 3) and Eq. (3) for the rotated channel; both findings are consistent with data gathered on hex-YMnO$_3$ at a temperature 10 K at the space group forbidden reflection (0, 0, 1) [14]. Moreover, intensities measured at L$_{2, 3}$ absorption edges are equal, to a good approximation [14]. The observation implies that Mn orbital states are negligible in the dipole $\langle T^1_\zeta \rangle$ if intensities are enhanced by an E1-E1 event, and appropriate sum rules are examined in Appendix B. Dirac multipoles contribute to diffraction enhanced by a parity-odd E1-E2 event. As with parity-even E1-E1, unrotated amplitudes are zero. Eq. (4) for the rotated amplitude is proportional to the diagonal component of the Dirac quadrupole and independent of the azimuthal angle. E1-E1 and E1-E2 amplitudes differ with respect to phase and the Bragg angle. Non-magnetic polar multipoles that are signatures of the polar character of the space group have zero amplitudes for (0, 0, *l*) with odd *l*. Looking ahead to future experiments, an anapole depicted in Fig. 2 parallel to the tertiary symmetry direction [1, 2, 0] contributes to all E1-E2 amplitudes with a reflection vector (*h*, 0, *l*) with odd *l*, e.g., Eq. (6).

Magnetic neutron diffraction by space group P6$_3$'cm' is discussed in Appendix C using (*h*, 0, *l*) with odd *l*. Diffraction by axial multipoles includes a dipole parallel to the ξ axis [1, 0, 0] that is likely the dominant feature at small wavevectors κ ≈ 0, cf. Fig. 4. A quadrupole due to the correlation of the spin anapole (**S** × **n**) and orbital **n** operators enters at around κ ≈ 6 Å$^{-1}$. Notable features of diffraction by Dirac multipoles are contributions by quadrupoles $\langle H^2 \rangle \propto$

[($h_1$) ⟨{**S** ⊗ **n**}$^2$⟩], with the radial integral ($h_1$) depicted in Fig. 4. Exact same quadrupoles explain neutron diffraction from high-$T_c$ compounds Hg1201 and YBCO [23, 24].

Resonant x-ray diffraction presented by the magnetic structure P6$_3$' is discussed in Section IV; it is in the history of hex-YMnO$_3$ [4, 5]. Manganese ions occupy sites in P6$_3$' that have no symmetry. Even so, patterns for P6$_3$' and P6$_3$'cm' are identical for reflections (0, 0, $l$) with odd $l$. Scope to observe the reduced symmetry in P6$_3$' exists in reflections ($h$, 0, $l$) and odd $l$. To this end, we give the E1-E2 unrotated amplitude Eq. (8).

Ramakrishnan *et al.* find an interpretation of their resonant x-ray diffraction data in terms of a simulation of the electronic structure of hex-YMnO$_3$ based on the magnetic space group P6$_3$c'm' (No. 185.201) that belongs to the magnetic crystal class 6m'm' [14, 15]. A comparison of P6$_3$c'm' and P6$_3$'cm' reveals that Mn ions have identical spatial coordinates and site symmetries, and the piezomagnetic effect is allowed in 6m'm' and 6'mm'. Ferromagnetism and a linear magnetoelectric effect are permitted in P6$_3$c'm' and forbidden in P6$_3$'cm'. There is no experimental evidence that these properties are displayed by hex-YMnO$_3$, to the best of our knowledge. Turning to the electronic structure factor for reflections (0, 0, $l$), the time signature $\sigma_\theta$ is not explicit for P6$_3$c'm', whereas our result Eq. (2) for P6$_3$'cm' possesses the diffraction condition $[\sigma_\theta(-1)^{l+Q}] = +1$ that is highly influential in scattering amplitudes. Observed spectral line shapes at L$_{2,3}$ absorption edges are different at different temperatures [14]. The interpretation offered appeals to interference of axial and Dirac scattering amplitudes, using a Dirac octupole Re. ⟨G$^3_{+3}$⟩. Axial and Dirac amplitudes Eqs. (3) and (4) for P6$_3$'cm' do not interfere, because of a 90º degree phase shift, and site symmetry requires Re. ⟨G$^3_{+3}$⟩ = 0 in both P6$_3$'cm' and P6$_3$c'm'.

**ACKNOWLEDGEMENTS**. Correspondence with Dr Y. Joly and Dr U. Staub was essential to unmasking different magnetic symmetries in this work and Ref. [14]. Dr Valerio Scagnoli provided Fig. 2. Professor G. van der Laan performed calculations of atomic radial integrals depicted in Fig. 4. Valuable guidance on the use of crystallography came from Dr K. S. Knight.

### APPENDIX A: ELECTRONIC STRUCTURE FACTOR

Electronic structure factor Eq. (1). Let $\Phi = \sigma_\theta (-1)^{l+Q}$, with $\Phi = -(-1)^{l+Q}$ for Dirac multipoles, and $\Phi = (-1)^{l+K+Q}$ for E1-E1 and E2-E2. A reflection vector ($h$, $k$, $l$) and sites 6c in symmetry P6$_3$'cm',

$$\Psi^K_Q(6c) = \langle O^K_Q \rangle [\alpha + \alpha^* \Phi + \exp(i2\pi Q/3) \{\beta + \beta^* \Phi\} \quad (A1)$$

$$+ \exp(-i2\pi Q/3) \Phi \{\alpha\beta + \alpha^*\beta^* \Phi\}].$$

Here, $\alpha = \exp(i2\pi hx)$ and $\beta = \exp(i2\pi kx)$, with x ≈ 0.342 [6]. As the material is ferroelectric (polarization along c axis), the structure factor is arbitrary to within a phase that we set equal to unity (z = 0).

## APPENDIX B: SUM RULES

Subsequent results are L edge RMEs of electronic multipoles $(\chi\|O^K\|\chi')$, where $\chi$ is a composite label for necessary quantum numbers excluding projections Q. An expectation value $\langle O^K_Q \rangle$ is a sum of RMEs each multiplied by a 3j-symbol that is the sole bearer of Q (Wigner-Eckart Theorem). Furthermore, an RME is written in terms of standard unit tensors $W^{(a, b)K}$ with spin variable a = 0 or 1, an orbital variable b, and even (a + b + K) [20]. There are equivalent operators for unit tensors that expose their physical content. Equivalent operators for dipoles are $W^{(1, 0)1} \propto \mathbf{S}$, $W^{(0, 1)1} \propto \mathbf{L}$, and $W^{(1, 2)1} \propto [\mathbf{S}(\mathbf{R}\cdot\mathbf{R}) - 3\mathbf{R}(\mathbf{S}\cdot\mathbf{R})]$ where $\mathbf{R}$ is the position operator conjugate to $\mathbf{L}$. For the E1-E1 dipole $\langle \mathbf{T}^1 \rangle$,

$$A = -\sqrt{(1/30)}\, W^{(0, 1)1},\quad B = 1/(3\sqrt{15})\,[W^{(1, 0)1} - \sqrt{35}\, W^{(1, 2)1}]. \tag{B1}$$

The coefficients are correct for L edges and d-like valence states. Turning to RMEs for E2-E2 and 2p → 4p, A is the same as for E1-E1 with $\langle \mathbf{T}^1 \rangle$ an expectation value of Mn 4p states. Orbital B is the sum of a dipole and an octupole $W^{(1, 2)3}$ that determines $\langle T^3_0 \rangle$ in $(\pi'\sigma)_{22}$. For an E2-E2 event,

$$B = -1/(3\sqrt{5})\,[W^{(1, 0)1} - W^{(1, 2)1}] - \sqrt{(2/15)}\, W^{(1, 2)3}, \tag{B2}$$

with $\langle T^3_0 \rangle \propto \langle W^{(1, 2)3} \rangle_0 \propto \langle [5S_\zeta R_\zeta^2 - 2S_\zeta (\mathbf{R}\cdot\mathbf{R}) - R_\zeta (\mathbf{S}\cdot\mathbf{R})] \rangle$.

RMEs for an E1-E2 event are listed by Lovesey and Balcar [35].

## APPENDIX C: MAGNETIC NEUTRON DIFFRACTION

Magnetic multipoles in neutron diffraction depend on the magnitude of the reflection vector, $\kappa$ [22, 29, 30]. An axial dipole $\langle \mathbf{t}^1 \rangle$ contains $\langle j_0(\kappa) \rangle$ and $\langle j_2(\kappa) \rangle$ that are averages of spherical Bessel functions of order 0 and 2 with respect to the radial density of the open shell. By definition, $\langle j_0(0) \rangle = 1$ and $\langle j_2(0) \rangle = 0$, and results in Fig. 4 are illustrative of their dependence on $\kappa$ [31]. A guide to the transition-metal dipole,

$$\langle \mathbf{t}^1 \rangle \approx (\langle \boldsymbol{\mu} \rangle/3)\,[\langle j_0(\kappa) \rangle + \langle j_2(\kappa) \rangle\,(g - 2)/g], \tag{C1}$$

is often used [29, 32]. Here, the magnetic moment $\langle \boldsymbol{\mu} \rangle = g \langle \mathbf{S} \rangle$ and the orbital moment $\langle \mathbf{L} \rangle = [(g - 2) \langle \mathbf{S} \rangle]$. The coefficient of $\langle \mathbf{L} \rangle$ in Eq. (C1) is approximate, while $\langle \mathbf{T}^1 \rangle = (1/3) \langle 2\mathbf{S} + \mathbf{L} \rangle$ for $\kappa \to 0$ is an exact result. Multipoles with even K arise from electron spin and spatial degrees of freedom, e.g., $\langle t^2_0 \rangle \propto [\langle j_2(\kappa) \rangle \langle (\mathbf{S} \times \mathbf{n})_0 n_0 \rangle]$ with $\mathbf{n} = \mathbf{R}/R$, and a maximum of $\langle j_2(\kappa) \rangle$ around $\kappa \approx 6$ Å$^{-1}$ [22]. Axial multipoles have ranks K = 1-5 for d-type ions. Intensity of a magnetic Bragg spot = $|\langle \mathbf{Q}_\perp \rangle|^2$ when the neutron beam is unpolarized. In more detail, $\langle \mathbf{Q}_\perp \rangle^{(\pm)} = [\mathbf{e} \times (\langle \mathbf{Q} \rangle^{(\pm)} \times \mathbf{e})]$ with a unit vector $\mathbf{e} = \boldsymbol{\kappa}/\kappa$, and superscripts refer to axial (+) and Dirac (−) multipoles. The intermediate amplitude $\langle \mathbf{Q} \rangle^{(+)}$ is proportional to the axial magnetic moment $\langle \boldsymbol{\mu} \rangle$ in the forward direction of scattering, with $\langle \mathbf{Q} \rangle^{(+)} = \langle \boldsymbol{\mu} \rangle/2$ for $\kappa = 0$, while the dipole in $\langle \mathbf{Q} \rangle^{(-)}$ can be related to anapoles [22, 33]. In place of Eq. (C1), the Dirac dipole $\langle \mathbf{d} \rangle$ depends on three radial integrals,

$$\langle \mathbf{d} \rangle = (1/2) \, [ \, i(g_1) \langle \mathbf{n} \rangle + 3 \, (h_1) \langle \mathbf{S} \times \mathbf{n} \rangle - (j_0) \langle \boldsymbol{\Omega} \rangle]. \qquad (C2)$$

Radial integrals $(g_1)$ and $(j_0)$ depicted in Fig. 4 diverge in the forward direction of scattering. Not so for $(h_1)$ that accompanies a spin anapole $\langle \mathbf{S} \times \mathbf{n} \rangle$. It is also the $\kappa$-dependence of the Dirac quadrupole $\langle \mathbf{H}^2 \rangle \propto [(h_1) \langle \{\mathbf{S} \otimes \mathbf{n}\}^2 \rangle]$ observed in neutron diffraction from high-$T_c$ compounds Hg1201 and YBCO [23, 24]. Returning to Eq. (C2), $\langle \boldsymbol{\Omega} \rangle = [\langle \mathbf{L} \times \mathbf{n} \rangle - \langle \mathbf{n} \times \mathbf{L} \rangle]$ is an orbital anapole (toroidal dipole) depicted in Fig. 2.

Neutron scattering amplitudes for $\kappa = (0, 0, l)$ with odd $l$ are identically zero, for both axial and Dirac multipoles. Results for $(h, 0, l)$ and odd $l$ are complicated and we limit results to low-order multipoles. Amplitudes have a common factor $[2i \sin(2\pi h x)]$ that we omit from the following results. With $\mathbf{e} = (e_\xi, e_\eta, e_\zeta) = (h\sqrt{3}, h, rl\sqrt{3})/Z$ and $\mathbf{e} \cdot \mathbf{e} = 1$,

$$\langle Q_\xi \rangle^{(+)} \approx (9/4) \langle t^1_\xi \rangle + (3/2) \sqrt{3} \, e_\zeta^2 \, \langle t^2_{+1} \rangle'' + (3/16) \sqrt{21} \, (3 - 7 \, e_\zeta^2) \langle t^3_{+1} \rangle',$$

$$\langle Q_\eta \rangle^{(+)} \approx (3/4) \sqrt{3} \langle t^1_\xi \rangle + (3/2) \, e_\zeta^2 \, \langle t^2_{+1} \rangle'' + (3/4) \sqrt{7} \, (7 \, e_\eta^2 - 1) \langle t^3_{+1} \rangle',$$

$$\langle Q_\zeta \rangle^{(+)} \approx - 6 \, e_\eta \, e_\zeta \, (\langle t^2_{+1} \rangle'' + \sqrt{7} \, \langle t^3_{+1} \rangle'). \qquad (C3)$$

The intensity for $(h, 0, l)$ with odd $l$,

$$|\langle \mathbf{Q}_\perp \rangle^{(+)}|^2 \approx (9/4)\sqrt{3} \, e_\zeta^2 \, \langle t^1_\xi \rangle \, [\sqrt{3} \langle t^1_\xi \rangle + 16 \, e_\eta^2 \, \langle t^2_{+1} \rangle''], \qquad (C4)$$

to a good approximation at a level of dipoles and quadrupoles. The equivalent intensity for diffraction by the Mn anapole and Dirac quadrupole $\langle H^2_{+1} \rangle'$ is,

$$|\langle \mathbf{Q}_\perp \rangle^{(-)}|^2 \approx 3 \, (2e_\zeta^2 + 1) \, [\langle d_\eta \rangle + (3/\sqrt{5}) \, (2e_\zeta^2 - 1) \langle H^2_{+1} \rangle']^2. \qquad (C5)$$

Both intensities vanish for $h = 0$ by virtue of a common factor $[\sin(2\pi h x)]^2$ not shown explicitly.

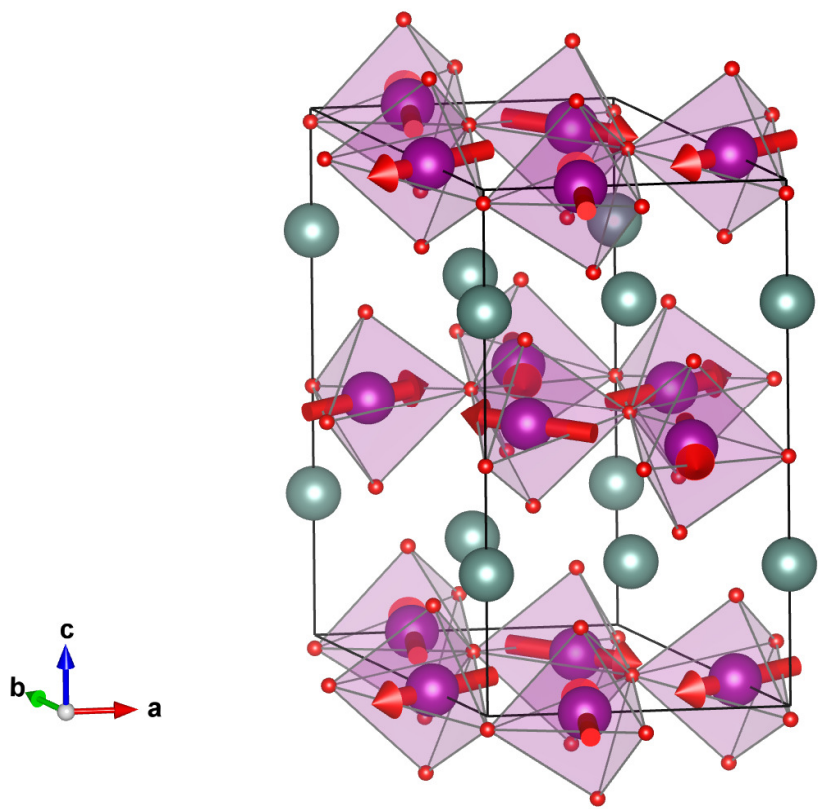

FIG. 1. Configuration of Mn axial dipoles in the (ab) plane of hex-YMnO$_3$. Oxygen (red) and Y (green) [1, 5, 6]. Reproduced from MAGNDATA [17].

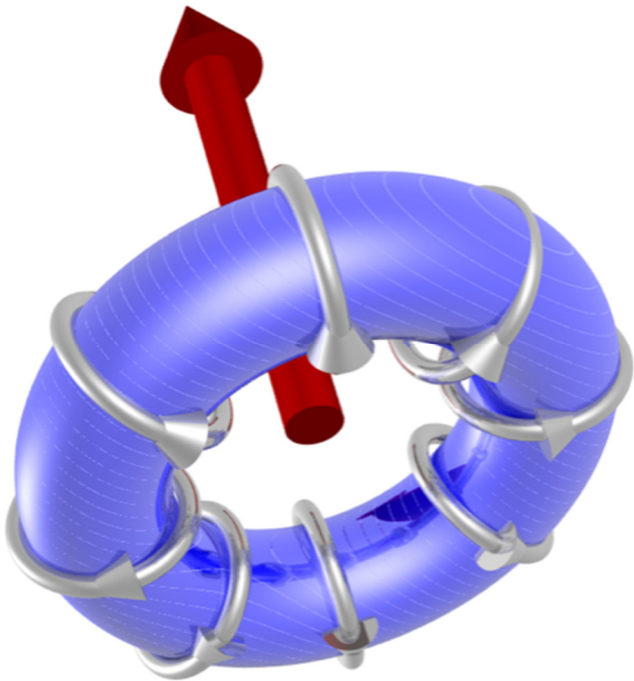

FIG. 2. Depiction of a toroidal dipole, also known as an anapole. Figure prepared by V. Scagnoli [21].

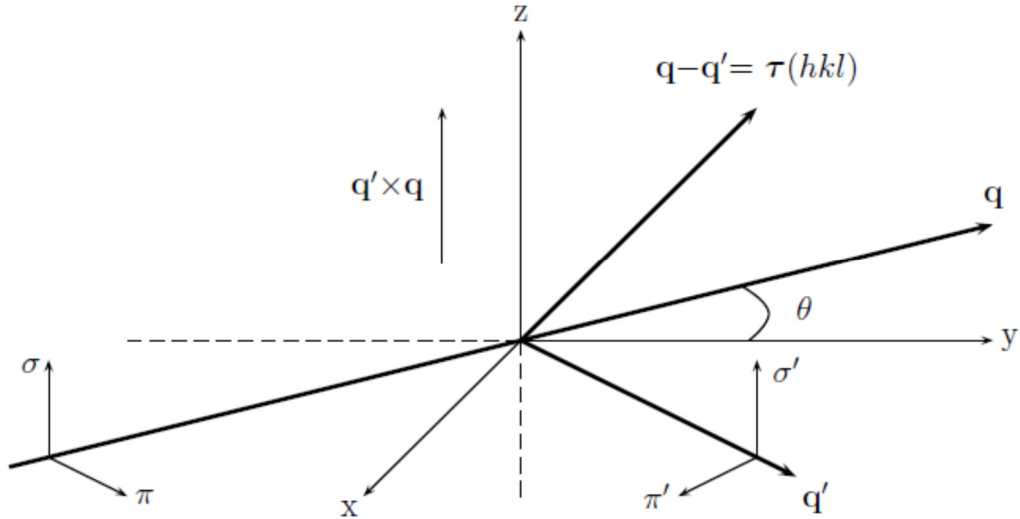

FIG. 3. Primary (σ, π) and secondary (σ', π') states of polarization. Corresponding wavevectors **q** and **q'** subtend an angle 2θ. The Bragg condition for diffraction is met when **q** − **q'** coincides with a vector τ($h$, $k$, $l$) of the reciprocal lattice. Crystal vectors **a**, **b***, and **c** that define local axes (ξ, η, ζ) and the depicted Cartesian (x, y, z) coincide in the nominal setting of the crystal.

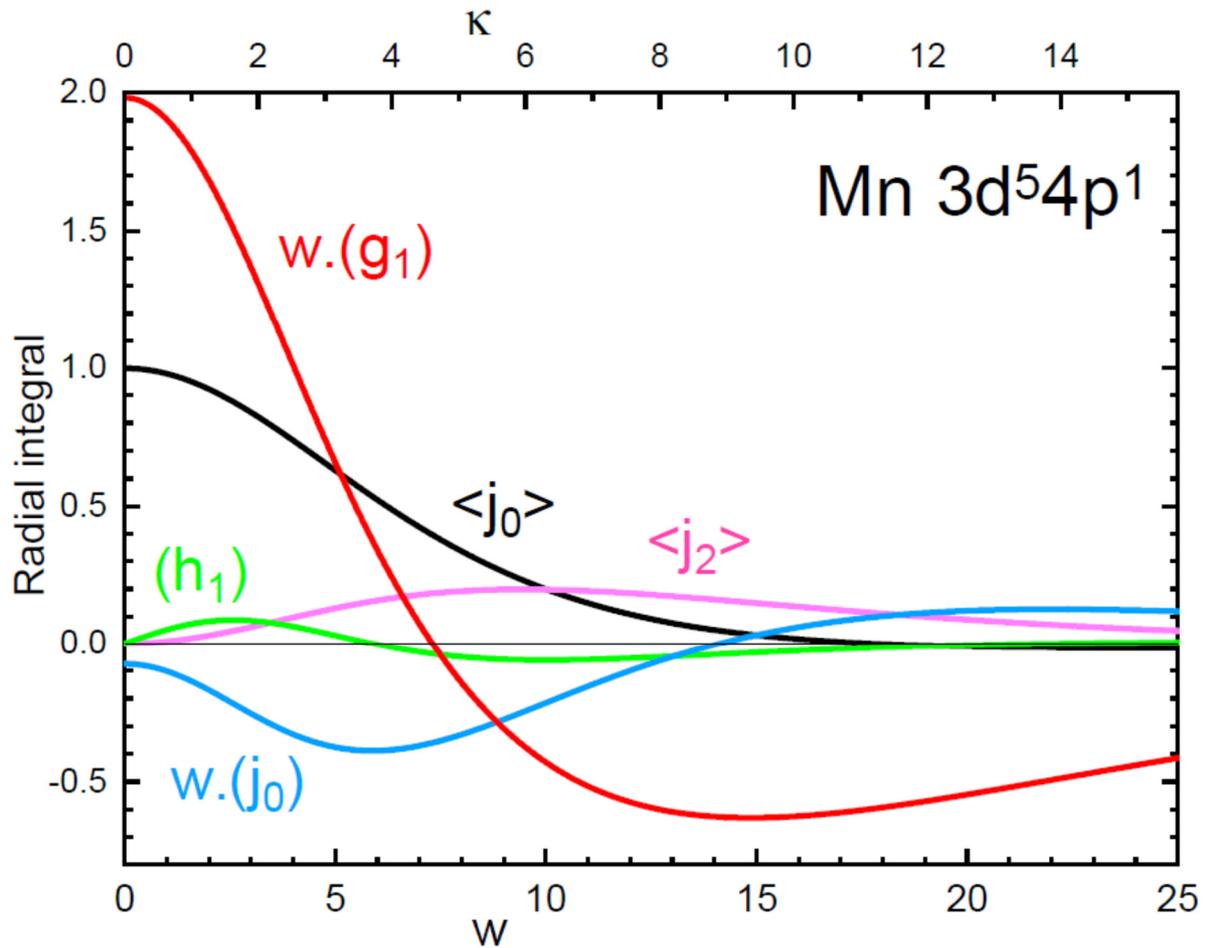

FIG. 4. Radial integrals for $Mn^{2+}$ ($3d^5$) displayed as a function of the magnitude of the reflection vector κ = 4π$s$ with $s$ = sin(θ)/λ (Å$^{-1}$), Bragg angle θ and neutron wavelength λ. Also, a dimensionless variable $w = 3a_o\kappa$ where $a_o$ is the Bohr radius. Blue and purple lines are standard radial integrals $\langle j_0(\kappa) \rangle$ and $\langle j_2(\kappa) \rangle$ that occur in the axial dipole Eq. (C1). Red, green and blue curves are radial integrals in the polar dipole Eq. (C2). Two integrals $(g_1)$ and $(j_0)$ diverge in the forward direction of scattering, and quantities $w(g_1)$ and $w(j_0)$ are displayed for this reason. Calculations, performed with Cowan's atomic code [34], and figure made by G. van der Laan.